\documentclass[onecolumn,doublespacing]{gji}
\usepackage{graphicx} 
\usepackage{amsmath}
\usepackage{amssymb}

\newcommand*\dd{\mathop{}\!\mathrm{d}}

\makeatletter
\let\bibhang\@undefined
\makeatother

\usepackage[round]{natbib}

\title[Exact seismograms in self-gravitating spherically symmetric Earth]
  {The direct spectral element method for the calculation of synthetic seismograms in self-gravitating, spherically symmetric planets}
\author[A.D.C. Myhill and D. Al-Attar]
  {A.D.C. Myhill$^1$, D. Al-Attar$^1$ \\
  $^1$ Bullard Laboratories, University of Cambridge, Madingley Road,
   Cambridge CB3 0EZ, UK 
  }
\date{Received 2026 March 6}
\pagerange{\pageref{firstpage}--\pageref{lastpage}}
\pubyear{2026}

\begin{document}
\label{firstpage}
\maketitle

\begin{summary}
This paper describes the implementation of the direct solution method (DSM) using radial spectral 
elements for the calculation of synthetic seismograms in self-gravitating, spherically symmetric, 
non-rotating, anelastic, and transversely isotropic Earth models. In contrast to 
previous implementations of the DSM that used a potential 
formulation within fluid regions, 
we use a displacement formulation throughout. It is this feature that allows us to
extend the DSM to account fully for self-gravitation along with arbitrary fluid stratification. 
Our code, \texttt{DSpecM1D}, is benchmarked against 
the normal mode summation code \texttt{MINEOS} as well as the direct radial integration code \texttt{YSpec}. Agreement between
 the codes is excellent for both elastic and anelastic models. 
\end{summary}


\section{Introduction}

\subsection{Context and motivation}

One of the most important features of long-period free oscillation spectra is their direct sensitivity to density, 
through the effect of self-gravity~\citep[e.g][]{dziewonski1981preliminary,Dahlen_Tromp_1998,ishii1999normal}.
While the Earth's spherically symmetric density structure is well-established, basic questions remain about lateral density variations within the mantle, including the nature of the Large Low Velocity Provinces \citep[e.g][]{ishii1999normal,davies2015thermally,lau2017tidal}.
In addressing such questions, free oscillation studies have and will continue to play a major role. Progress is, however, currently held back through difficulties in accurately and efficiently modelling free oscillation spectra in laterally heterogeneous Earth models. Developing a new open source code for this purpose is the principal motivation for this study. 
 
In comparison to seismological modelling at shorter periods, free oscillation work is complicated by 
the need to account fully for the effects of self-gravitation within the dynamics, 
while the influence of rotation and non-hydrostatic pre-stress also become more important
\citep[e.g.][]{woodhouse1978effect,valette1986influence,Dahlen_Tromp_1998}.
Despite progress in the incorporation of self-gravitation into 
the 3D spectral element method \citep[][]{chaljub2004spectral,van2021modelling,gharti2023spectral}, 
normal mode coupling remains the only practical method  for performing such 
calculations \citep[e.g.][]{woodhouse1980coupling,deuss2001theoretical,al2012calculation,
yang2015synthetic,akbarashrafi2018exact,adourian2024adjoint}. Indeed, the need within 
free oscillation studies to consider time series of hundreds of hours 
in length is likely to render time-domain 3D spectral element methods
prohibitively expensive for the foreseeable future. While frequency-domain 3D spectral element formulations may offer a viable route forward, they remain to be developed, and their relative costs established.

Normal mode coupling is based on a Galerkin formulation of the equations of motion, using, as the basis functions, a truncated set of the eigenfunctions of a spherically symmetric reference model. Current implementations of mode coupling theory account for the effect of density perturbations and boundary topography only up to first-order accuracy, but these limitations could be overcome using the particle relabelling methods introduced by \citet{al2016particle} and \citet{al2018hamilton} that builds on an idea of \citet{woodhouse1976rayleigh}. There remains, however, a subtle and potentially important limitation of normal mode coupling that, to our knowledge, has not been discussed in the literature. Consider, for simplicity, a spherical reference model, and an associated laterally heterogeneous model, with no topography on any of the boundaries. If we calculate a set of eigenfunctions within the reference model, then their tractions satisfy appropriate boundary and continuity conditions within the reference model. A linear combination of such functions cannot, in general, satisfy the correct boundary conditions within the laterally heterogeneous model because the tractions depend on the laterally heterogeneous elastic moduli at the boundaries. This suggests that numerical solutions obtained using normal mode coupling cannot converge to the true solution as the number of basis functions is increased. In short, the eigenfunction bases used within normal mode coupling have missing degrees of freedom associated with internal and external boundaries. In a separate study we will describe a method for addressing this point within the context of coupling calculations by suitably augmenting the usual eigenfunction bases. Here, however, we seek a different approach that can overcome the above issue while retaining the main advantages of normal mode coupling. 

To proceed, we retain the Galerkin formulation that is common to both mode coupling and spectral element formulations. For the basis functions, however, we plan to use an intermediate choice between the reference eigenfunctions in coupling calculations and a full 3D spectral element discretisation; fields are discretised radially using a 1D spectral element basis, and angularly with generalised spherical harmonic expansions. Such an approach has been applied by \cite{maitra2019non} and \cite{myhill2025forward} in combination with the particle relabelling method for the solution of Poisson's equation within aspherical planets. Within this method, the necessary linear system is solved using a preconditioned iterative method. Crucially, the matrix corresponding to the full problem is not assembled, but instead its action determined in a matrix-free manner using a hybrid pseudo-spectral/spectral element approach. To precondition this linear system, the solution of Poisson's equation within a spherically symmetric reference model is used\textemdash obtained efficiently by solving a sequence of 1D spectral element problems at each order and degree. A key aspect of this method is that the cost of the solution scales with the complexity of the laterally heterogeneous model. This property is shared by modern iterative approaches to normal mode coupling \citep[e.g.][]{al2012calculation} and by the AxiSem3D method \citep[e.g.][]{leng2016efficient,leng2019axisem3d}. These ideas all extend naturally to the frequency-domain form of the equations governing seismic wave propagation within self-gravitating Earth models. To implement the method, however, a key step is the development of the analogous ``spherical Earth preconditioner'' and that is the aim of this paper.

\subsection{Aims and contributions}

As discussed, our aim is the development of an efficient numerical method for solving the frequency-domain 
equations of motion within self-gravitating, spherically symmetric Earth models. As is well known, 
through spherical harmonic expansions, this problem decouples into systems of  radial ordinary 
differential equations for each degree and order \citep[e.g.][]{Dahlen_Tromp_1998}. There 
are two existing methods for solving the resulting equations in the frequency domain, 
but neither are well-suited for our planned application. The first is normal mode summation, and the 
second is the variant of the direct radial integration method by \cite{al2008calculation}
that includes self-gravitation. Normal mode summation provides solutions to the 
frequency domain equations by first solving the associated eigenvalue problem, but this latter
problem takes substantially more effort than solving the forced equations 
of motion at the desired frequencies. The direct radial integration method was designed precisely to avoid this issue,
but it does not readily provide for the solution throughout the model, and that is what is needed within
the preconditioning steps outlined above. 

Following \cite{maitra2019non} and \cite{myhill2025forward}, the ideal approach would instead be the 
solution of the spherically symmetric problem on the same radial mesh to be used 
within the discretisation of the laterally heterogeneous problem. Doing this 
amounts to an extension of the direct solution method (DSM) introduced by \cite{geller1990new}
and subsequently developed in a series of papers \citep[e.g.][]{geller1994computation,cummins1994dsm,
cummins1994dsm2,cummins1997computation}. Our code \texttt{DSpecM1D}, (Direct Spectral Element Method 1D), significantly extends the DSM for spherically 
symmetric models by incorporating self-gravitation, and allowing for arbitrary stratification within fluid regions. 
The method also allows for the exact implementation of  general linear viscoelastic rheologies with transversely isotropic parameters.
Our numerical implementation is based on  a high-order spectral element discretisation that increases both accuracy and efficiency
compared to low-order finite elements, with this approach being  equivalent to that  used by  by~\citet{kemper2022spectral}  in 
the solution of the associated eigenvalue problem.   
\texttt{DSpecM1D} is benchmarked against the  normal-mode summation code \texttt{MINEOS} and the direct radial integration code \texttt{YSpec}~\citep{mineos,al2008calculation}. 
While  the development of this code has been motivated by ongoing work related to free oscillation seismology, the new method is
of interest in its own right and has potential applications within higher frequency seismological modelling. 

To conclude the introduction, we note that this paper is intended be a relatively concise discussion. We do not dwell on the details of 
standard reductions of the equation of motion, or numerical techniques which are well established in seismology. In the main text we focus on the novel 
aspects of our implementation and their implications, especially the physical insights they bring. Further details of the numerics are provided in the 
appendices. For an introduction to the DSM we direct the reader to~\citet{geller1994computation} as well as the later papers by~\citet{cummins1994dsm,cummins1994dsm2} 
whilst for details on spectral element methods we recommend reading~\citet{komatitsch1998spectral} or~\citet{igel2017computational}.

\section{Theory and numerics}
There are two distinct aspects to this problem which we must consider: the first is the physics that we wish to explore, and the second the numerical solution of the equations of motion. In this respect we first discuss the equations of motion, the Earth models we consider and the phenomena that arise due to the fluid outer core. The discussion then turns to the numerical solution of the equations of motion.

\subsection{Equations of motion}
The non-rotating Earth model is assumed to occupy a region $M$, with external boundary $\partial M$. The union of all solid and fluid regions are respectively denoted $M_S$ and $M_F$. The union 
of all internal boundaries is denoted by $\Sigma$ and individual boundaries are denoted by $\Sigma_{SS}$ (solid-solid) and $\Sigma_{SF}$ (solid-fluid). The weak form of the viscoelastic
 wave equation in the frequency domain is given by
\begin{equation}
  -\omega^2 \mathcal{P}(\mathbf{u}', \mathbf{u}) + \mathcal{H}(\omega,\mathbf{u}', \mathbf{u})  = \mathcal{F}(\mathbf{u}'), 
\end{equation}
where $\mathcal{P}(\cdot,\cdot)$ and $\mathcal{H}(\cdot,\cdot)$ are the sesquilinear inertia and stiffness forms respectively, and $\mathcal{F}(\cdot)$ is a linear 
form associated with the seismic source or other force term. The displacement field is denoted by $\mathbf{u}$, $\mathbf{u}'$ is a test function, and the frequency is 
given by $\omega$. Within the weak formulation, both the trial and test functions must satisfy appropriate kinematic boundary and continuity conditions including the continuity of the displacement across solid-solid boundaries and the continuity of the normal component of the displacement across fluid-solid boundaries. All dynamic boundary conditions, such as the continuity of the traction across internal boundaries, are built directly into the weak formulation of the problem and need not be imposed explicitly.

\subsection{SNRATI Earth models}
We herein restrict the discussion to Earth models which are spherically symmetric, non-rotating, anelastic, transversely isotropic models (SNRATI). These models are described by a set of radially varying parameters: the density $\rho$, and the five frequency dependent Love parameters $A(\omega)$, $C(\omega)$, $N(\omega)$, $L(\omega)$, $F(\omega)$. This represents the most general, rotationally invariant Earth model~\citep{Dahlen_Tromp_1998}. We note that an isotropic model with bulk modulus $\kappa$ and shear modulus $\mu$ can be defined which is the closest, in a least-squares sense, to the SNRATI model. 

In an SNRATI Earth the spherical symmetry can be used to reduce the equations of motion to two sets of ordinary differential equations (ODEs), known as the spheroidal and toroidal equations, for which the independent variable is the radius. This procedure is described in greater detail in Appendix A1. In outline, the wavefield $\mathbf{u}$ can be expanded in terms of generalised spherical harmonics whose expansion coefficients vary radially. The equations for different degree and order spherical harmonic expansion coefficients are independent. Furthermore, the equations for a particular coefficient depend only on the degree and not on the order. The toroidal and spheroidal ODEs are given in Appendices A3 and A4. It should be emphasised  that the method (and indeed \texttt{DSpecM1D}) is not restricted to a particular class of spherically symmetric Earth model (e.g., those having the Earth-like structure of a solid inner core and a fluid outer core). \texttt{DSpecM1D} can be applied within fully general spherically symmetric models, such as purely solid planets, or planets with entirely fluid cores. However, we focus on the Earth in our discussion for definiteness.

\subsection{Fluid regions}
Within seismological applications, it is usual to regard the Earth's outer core and the oceans as inviscid fluids. The existence of such fluid regions presents a challenge within numerical modelling. To understand its implications, we consider a small parcel of fluid, which can be displaced. Restricting our discussion to radial models and radial displacement of the parcel, we can define the Brunt-V\"ais\"al\"a frequency (see e.g. Chapter 8 of~\citet{Dahlen_Tromp_1998}), given by 
\begin{equation}
  N^2 = -\frac{g\dot{\rho}}{\rho} - \frac{\rho g^2}{\kappa},
\end{equation}
where $\rho$ is the density, $g$ is the acceleration due to gravity, $\dot{\rho}$ is the vertical density gradient, and $\kappa$ is the bulk modulus (fluid regions are isotropic). If $N^2$ is positive, any radial displacement of a parcel of fluid will result in oscillation about the original position. If $N^2$ is negative, the equilibrium is unstable. In the case of a neutrally stratified fluid, $N^2 = 0$, and displacement will not result in oscillation. In spherically symmetric Earth models such as PREM \citep[][]{dziewonski1981preliminary}, the Brunt-V\"ais\"al\"a frequency is generally close to, but not exactly, zero. From a physical perspective, whilst we might expect that a vigorously convecting fluid will be close to a state of neutral stratification, this condition need not be met precisely. Indeed, it is not even certain that the entirety of the Earth's outer core is undergoing strong convection \citep[e.g.][]{helffrich2013causes}. 

It has been shown by \cite{valette1989spectre} that inclusion of fluid regions within a planet complicates its free oscillation spectrum significantly, resulting in the existence of an essential spectrum \citep[e.g.][]{schechter2001principles} in a bounded neighbourhood of the origin whose extent is, in the non-rotating case, determined by its Brunt-V\"ais\"al\"a frequency. Points within the essential spectrum need not be isolated, and the associated eigenfunctions can have very complicated forms within fluid regions whilst being near zero within solid parts of the model. This behaviour reflects the fact that within an inviscid fluid there is no resistance to shearing, and hence no minimum length-scale for shear motions at a given frequency. Within spherically symmetric models, it can be shown that the spectrum remains discrete, and isolated, except for at the origin which is an accumulation point. Moreover, as this accumulation point is approached, the radial eigenfunctions become more and more oscillatory.

\subsection{Galerkin discretisation and spectral element method}
The Galerkin method uses the same basis functions for the wavefield as the test functions. The basis that we choose to use is a one dimensional spectral element basis. We divide the domain $\mathcal{R}$ into a set of elements $\mathcal{R}_i \equiv [r_i,r_{i+1}]$. Within the $i$-th element we define a set of Gauss-Lobatto-Legendre (GLL) polynomials, these being Lagrange polynomials defined on the GLL quadrature nodes. We label the basis functions as $l_{\alpha}$ where $\alpha$ is a shorthand label, which points to both the element and the basis polynomial within that element. The ultimate result of this procedure is to transform the ODEs into a linear algebra problem at each frequency. In other words, we arrive at a linear system of the form 
 \begin{equation}
  [-\omega^2 \mathbf{P} + \mathbf{H}(\omega)] \mathbf{u}(\omega) = \mathbf{f}(\omega), \label{eqn:freqbaseeqn}.
\end{equation}
Continuity of displacement on solid-solid element boundaries is enforced by requiring the values on the upper and lower sides to be the same. On fluid-solid boundaries we allow tangential slip, as in~\citet{kemper2022spectral}.

The solution to eq.~\eqref{eqn:freqbaseeqn} is found at a set of frequencies $\{\omega_i\}$ and an inverse Fourier-Laplace (IFL) transform performed to find the time-domain solution. The frequencies $\{\omega_i\}$, at which eq.~\eqref{eqn:freqbaseeqn} are solved, have a small imaginary component, whose effect is removed after transforming to the time-domain by multiplying by an exponential function of time. This ensures the well-posedness of the IFL transform, as well as reducing aliasing in the time domain~\citep{geller1994computation,al2012calculation}. The matrix $[-\omega^2\mathbf{P} + \mathbf{H}(\omega) ]$ is symmetric but not Hermitian; sparse LU decomposition is performed at each frequency using the Eigen library~\citep*{eigenweb}. Finally, at a particular value of $l$ and $\omega$ there is a turning depth, below which the solution becomes evanescent~\citep{Dahlen_Tromp_1998,kawai2006complete}. For large $l$ and low frequencies this turning depth can be very shallow. Using JWKB theory \citep[e.g.][]{Dahlen_Tromp_1998}, we find a depth below which the classical approximation decays to less than some tolerance with respect to its value at the turning point. The matrix equations are truncated at this depth, effectively adding a free boundary condition~\citep{kawai2006complete}. This leads to very significant speed-ups, especially as the maximum frequency is increased.

\subsection{Numerical considerations}
As noted in the discussion on the theoretical aspects of the problem, the presence of undertones in the outer core presents challenges to numerical methods. To illustrate the problem we consider the calculation of the Earth's normal modes. When one tries to solve the eigenvalue problem associated with eq.~(\ref{eqn:freqbaseeqn}), one finds that the undertones are aliased into the seismic regime~\citep{buland1984computation,kemper2022spectral}. In finite element modelling, this problem has typically been addressed by employing a single potential representation within the outer core, which is only valid if the outer core is neutrally stratified~\citep{geller1994computation,komatitsch1998spectral}. A two-potential representation due to~\citet{chaljub2004spectral} has also been developed, which allows for non-neutral stratification, though it is not clear whether such an approach has practical benefits. We have found that when solving for the ground motion with realistic (seismic) forcing terms, at seismic frequencies, undertones are not excited to an appreciable extent. Consequently, we do not need to enforce or make any assumptions about the stratification of the outer core. 

Incorporating attenuation (and dispersion) requires modifying the equations of motion to account for anelastic effects. As we are using a frequency domain method this is quite straightforward\textemdash we simply need to allow the Love parameters to be frequency dependent. The inertia matrix $\mathbf{P}$ is frequency independent, whilst the stiffness matrix $\mathbf{H}$ needs to be re-evaluated at each frequency. The most computationally efficient way of doing this varies depending on the model of anelasticity. Although the exact implementation details may vary, frequency domain methods make its inclusion straightforward. An example using the constant-Q dispersion model is given in Appendix B2.

Finally, extensive use has been made of parallelisation. There are several ways to go about parallelisation, as the equations are independent for different degrees and different frequencies. We chose to parallelise over degree $l$, which we have found to be most efficient. It should be noted that in both the degree and frequency the equations of motion are ``embarrassingly'' parallel, as the equations at different frequencies and degrees are independent.

\section{Examples}
In this section we present benchmarks of \texttt{DSpecM1D}, consider its convergence, and explore the effect of the fluid outer core. For the purpose of benchmarking we compare against \texttt{MINEOS}, a normal-mode summation code~\citep{mineos}, and \texttt{YSpec}, a direct radial integration code~\citep{al2008calculation}. We take a similar philosophy to benchmarking to~\citet{mineos},~\citet{nissen2014axisem} and~\citet{gharti2023spectral}. In particular, we make visual comparisons of spectra and waveforms, and check the average differences between the codes. We define the simple percentage misfit function between two time-series $s_1(t)$ and $s_2(t)$ as
\begin{equation}
  \epsilon(t) = \frac{|s_1(t) - s_2(t)|}{s_{\text{max}}} \times 100.
\end{equation}
This is the same percentage misfit given by~\citet{gharti2023spectral}, as in that work we regularly quote the mean value of $\epsilon(t)$. Finally, it should be made clear that we do not seek to match the codes to numerical precision, as differences in the model representations and other implementation details preclude this. Nonetheless, discrepancies between the codes are significantly smaller than the expected difference with data. 

\subsection{Preliminary Reference Earth Model}
We simulate the propagation of seismic waves in the non-rotating, spherical, transversely isotropic PREM \citep[][]{dziewonski1981preliminary}. The ocean layer has been replaced by a crustal layer with $V_{\text{P}}=5800 \text{\,ms}^{-1}$, $V_{\text{S}}=3200 \text{ms}^{-1}$ and $\rho = 2600 \text{kg m}^{-3}$.  Attenuation is ignored in the first example.

\begin{figure*}
  \centering
  \includegraphics[width=\textwidth]{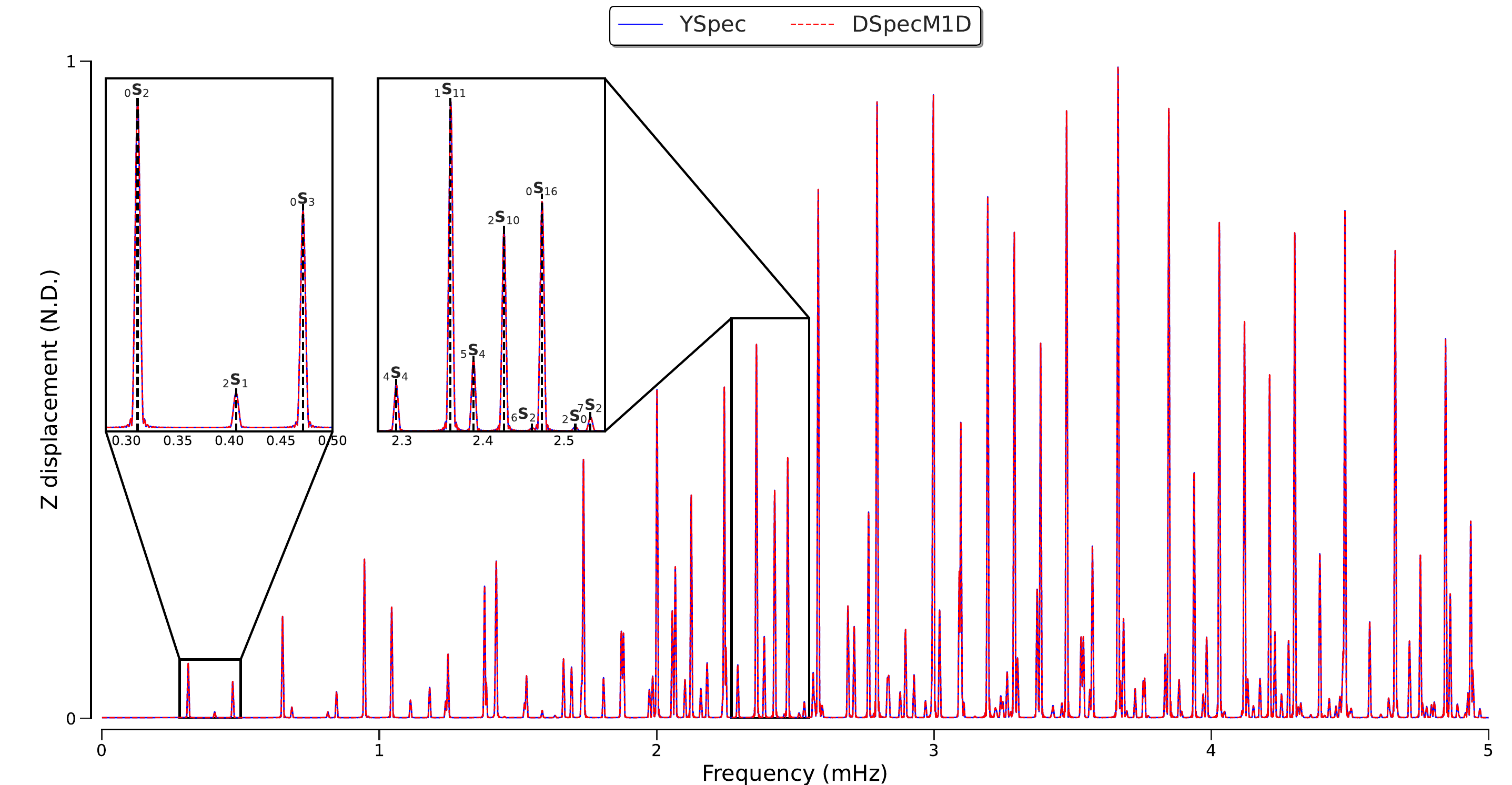}
  \caption{Comparison of vertical displacement spectra at location $(80^\circ \text{\,N}, 0^{\circ} \text{\,E})$ using the Bolivia 1994 source, for a 100 h time-series, up to maximum degree 100. The model is transversely isotropic, non-dispersive and non-attenuating PREM, without an ocean layer. \texttt{DSpecM1D} (red dashed line) is compared with the results from the direct radial integration method (\texttt{YSpec}, blue line). Between the two codes, the average relative misfit is 0.007\%, the maximum relative misfit is 0.32\%. In the inset axes we analyse more closely two portions of the spectra. The black dashed line in the inset axes indicates the frequency for the modes as computed by \texttt{MINEOS}, the corresponding modes are indicated above the peaks. }
  \label{fig:ex1a}
\end{figure*}

The first benchmark is for the 1994 deep-focus Bolivia Earthquake, at a depth of 647.1 km, location $(13.82^\circ\text{\,S}, 67.25^\circ\text{\,W})$. The receiver is placed at $(80.0^{\circ} \text{\,N}, 0^{\circ} \text{\,E})$. In Fig.~\ref{fig:ex1a} we compare the vertical displacement spectra found by \texttt{DSpecM1D} and \texttt{YSpec}, between $0$ and $5$\,mHz, for a 100\,h long time-series. The average difference between the two codes is on the order of $10^{-2}$ per cent. In addition, we have zoomed into two regions where we indicate the frequencies of the corresponding normal modes found using \texttt{MINEOS}. Agreement is excellent between all codes.

\begin{figure*}
  \centering
  \includegraphics[width=\textwidth]{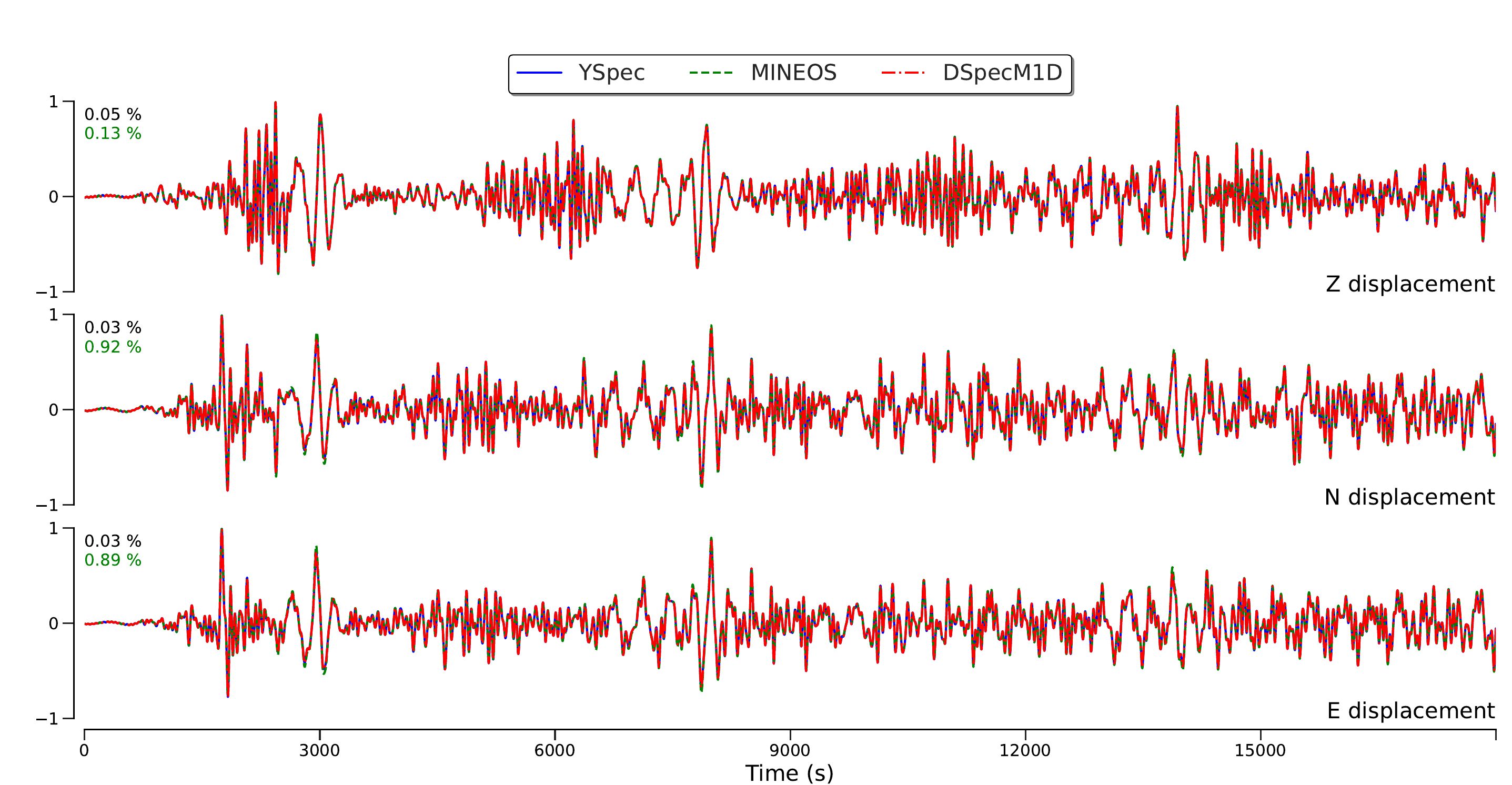}
  \caption{Comparison of displacement traces at $(80.0^{\circ} \text{\,N}, 0.0^{\circ} \text{\,E})$ for the Bolivia 1994 source, in a non-attenuating, non-dispersive, transversely isotropic PREM without an ocean layer. Calculated for all modes between 1.5 mHz and 21mHz, maximum degree 300. A half-cosine taper was applied with corner frequencies $(1.9,2.0)$ and $(20.0,20.1)$ mHz. The average relative differences between \texttt{DSpecM1D} and \texttt{YSpec} (\texttt{MINEOS}) are written on each plot in black (green), for each component respectively. The maximum relative difference (across all three components) was 0.4\% for \texttt{YSpec} and 7.3\% for \texttt{MINEOS}.}
  \label{fig:ex1b}  
\end{figure*}

In Fig.~\ref{fig:ex1b} we compare the displacement seismograms in all three channels (vertical, North and East) for a 5\,h time-series, filtered between 50 and 500\,s, at $(80.0^{\circ} \text{\,N}, 0^{\circ} \text{\,E})$ for the 1994 Bolivia Earthquake. All three codes give very similar results. The mean misfit between \texttt{DSpecM1D} and \texttt{YSpec} is significantly smaller than one per cent for all three channels, whilst it is around 1 per cent between \texttt{MINEOS} and \texttt{DSpecM1D}. 

\begin{figure*}
  \centering
  \includegraphics[width=\textwidth]{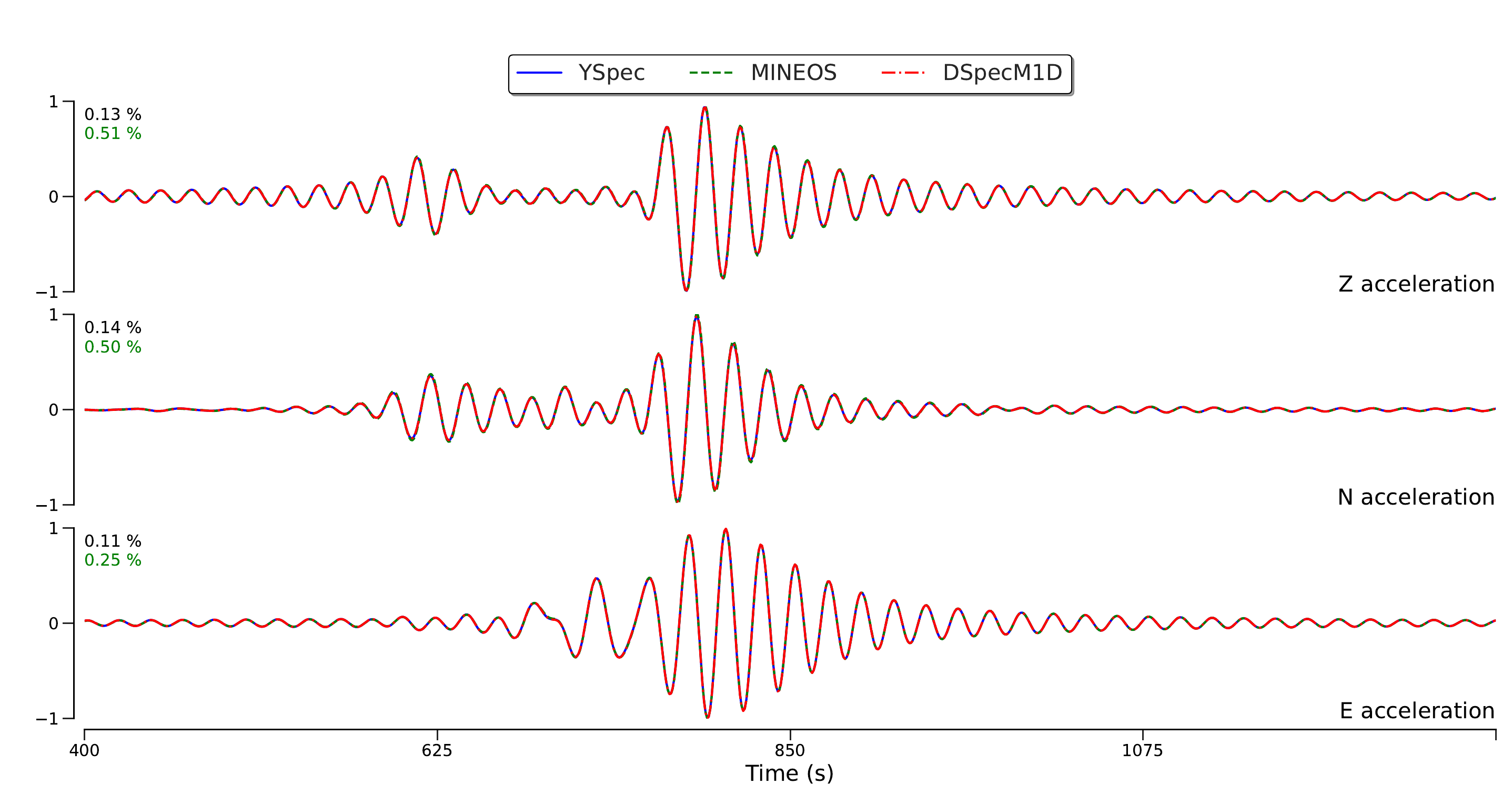}
  \caption{Comparison of acceleration traces at TLY station, located at $(51.68^{\circ} \text{\,N}, 103.64^{\circ} \text{\,E})$, for the China source, in an attenuating, dispersive, transversely isotropic PREM without an ocean layer. Calculated for all modes up to 51mHz, maximum degree 750. A half-cosine taper was applied with corner frequencies $(0.1,0.2)$ and $(50.0,50.5)$ mHz. The average relative differences between \texttt{DSpecM1D} and \texttt{YSpec} (\texttt{MINEOS}) are written on each plot in black (green), for each component respectively. The maximum relative difference (across all three components) was 1.5\% for \texttt{YSpec} and 3.3\% for \texttt{MINEOS}.}
  \label{fig:ex2}
\end{figure*}

The third benchmark is for an Earthquake in China, used as the benchmark for \texttt{MINEOS}, located at $(25.39^{\circ} \text{\,N}, 101.40^{\circ} \text{\,E})$, with a depth of 33\,km. The receiver is station TLY, location $(51.68^{\circ} \text{\,N}, 103.64^{\circ} \text{\,E})$. We include attenuation and dispersion and plot the three solutions for the acceleration in all three components in Fig.~\ref{fig:ex2}. The time window chosen corresponds to the first major arrival of surface waves. The seismograms are filtered between 20\,s and 5000\,s. Again, the differences are less than one per cent between the three codes. 

\subsection{Travel time benchmark}
\begin{figure*}
  \centering
  \includegraphics[width=\textwidth]{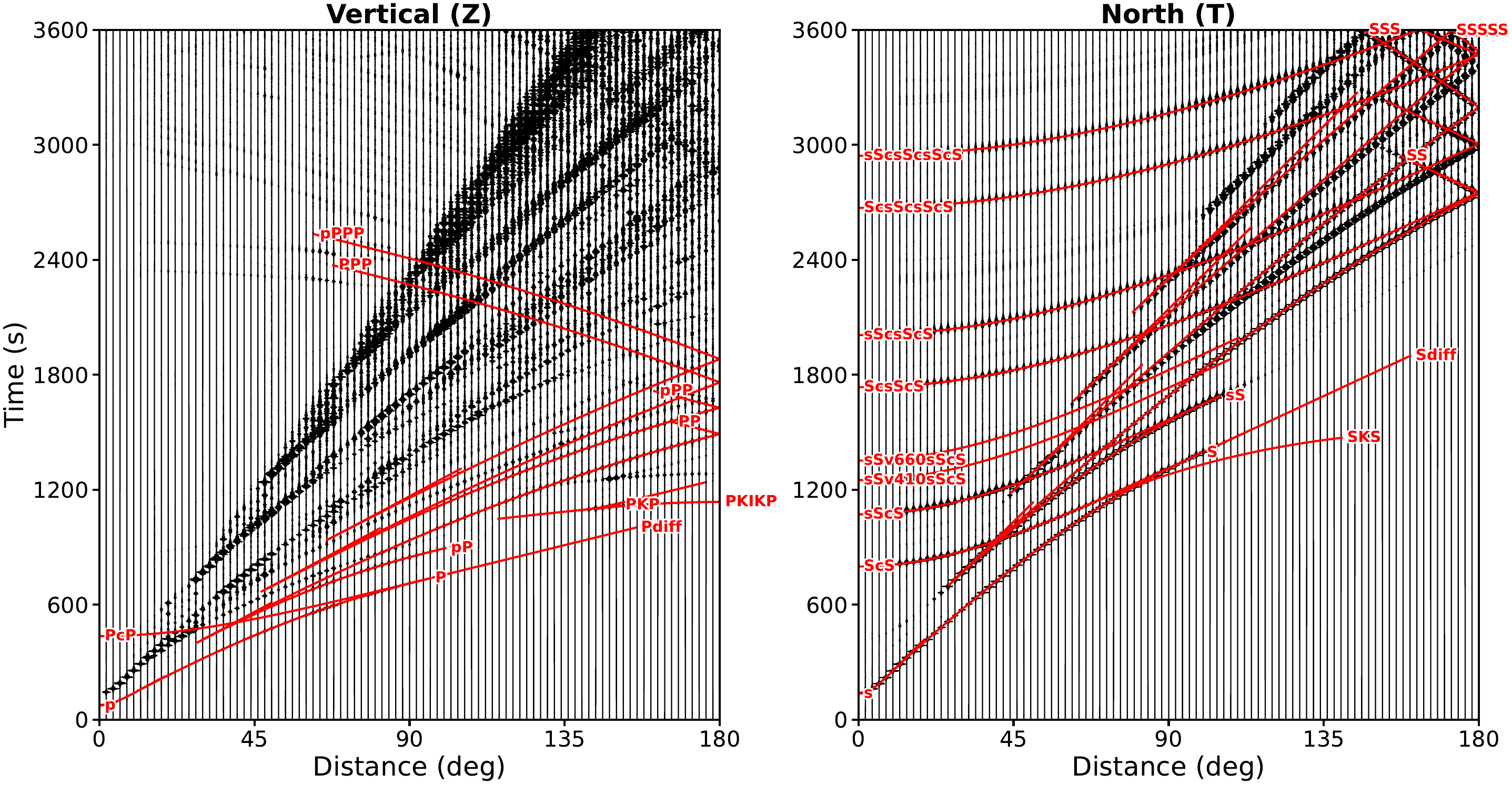}
  \caption{Record section for vertical and northward motion, up to 500 mHz, maximum degree 7500. The Earthquake source is the moment tensor for the Bolivia 1994 Earthquake, placed at $(0^{\circ} \text{\,N}, 0^{\circ} \text{\,E})$. A half-cosine taper applied with corner frequencies $(0.1,5.0)$ and $(450.0,500.0)$ mHz. Major seismic phase arrival times are given by the red lines, as calculated by TauP~\citep{crotwell1999taup}.}
  \label{fig:record_section}
\end{figure*}

As a further benchmark, and to demonstrate the versatility of \texttt{DSpecM1D} for higher frequency seismological applications, we compute a complete record section up to 2s period, at 91 equispaced receivers along the equator for a source with the same depth and moment tensor as the Bolivia Earthquake, but placed at $(0^\circ \text{\,N}, 0^\circ \text{\,E})$ (see Fig.~\ref{fig:record_section}). The vertical and northern components are illustrated. We use TauP to calculate the travel times of the different seismic phases shown~\citep{crotwell1999taup,crotwell_2025_16884103}. The motion was calculated within the frequency range $(0.1,500)$\,mHz. The total time-series shown is one hour long. 

\subsection{Convergence with element size}
\begin{figure*}
  \centering
  \includegraphics[width=\textwidth]{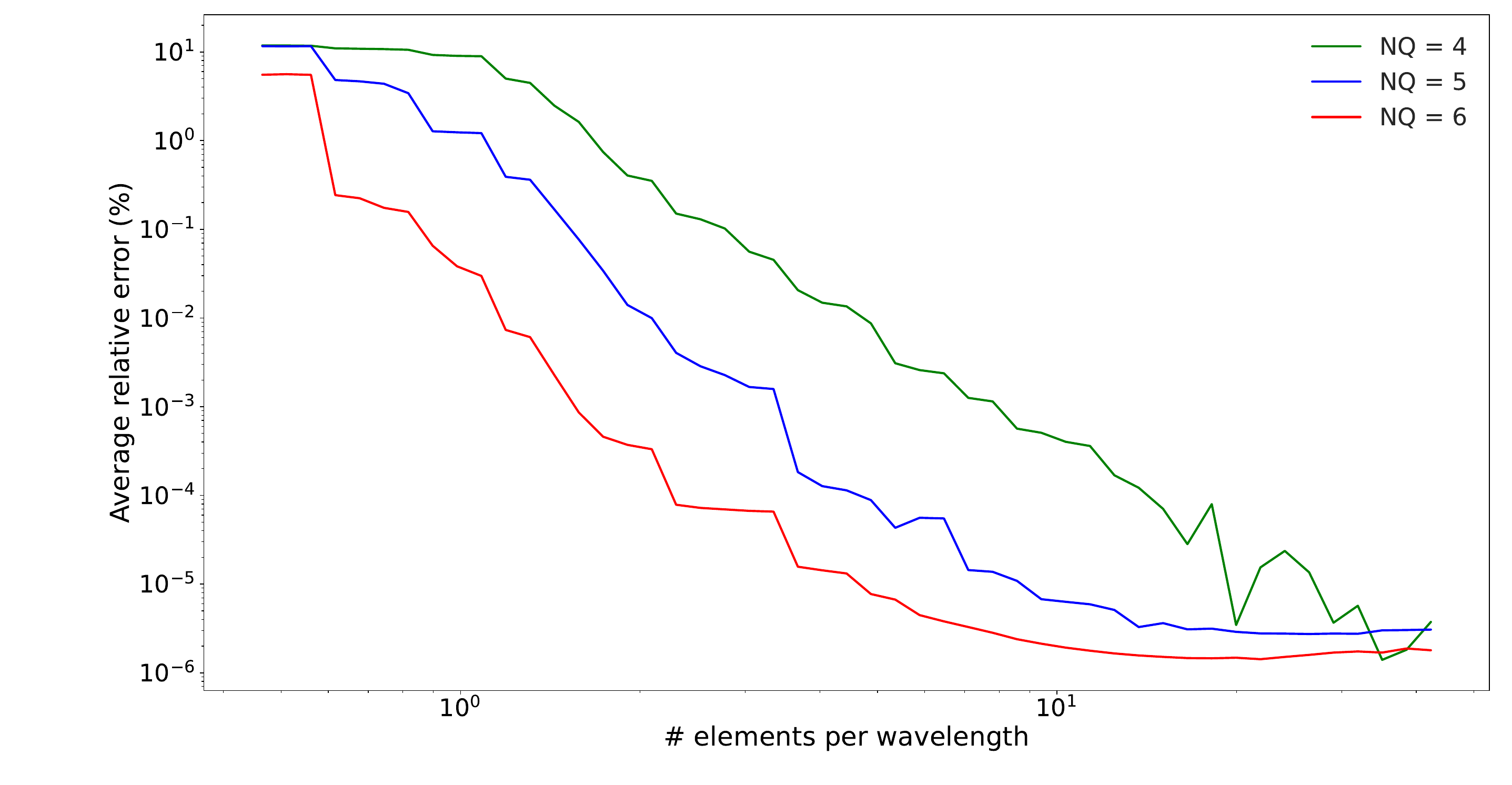}
  \caption{Average relative error computed in the time domain for a 500 minute time-series, compared to the exact solution as a function of element size. The time series is computed at station TLY for the Bolivia 1994 Earthquake. The relative error is calculated using the formula in the text. Maximum frequency is 50mHz, up to degree 750. The half-cosine taper was applied with corner frequencies $(0.1, 0.15)$, $(45.0, 50.0)$ mHz.}
  \label{fig:exconv}
\end{figure*}

One of the important computational considerations is the maximum size of elements within the mesh; the spectral element basis must be able to accurately represent the smallest wavelength oscillations. To quantify this we take a mesh with very small element size, with the resultant seismogram considered ``exact''. We then re-compute the seismogram with different maximum element sizes. The difference with the ``exact'' seismogram is then calculated as a function of the element size, as well as the number of points used in the spectral element method. The results of this are plotted in Fig.~\ref{fig:exconv}. Higher order methods have a faster convergence than lower order methods, as one expects~\citep{diaz1977collocation}. To achieve a relative error of 0.1\%, a 4 point SEM requires approximately three elements per minimum wavelength, whilst for a 6 point SEM, one element per wavelength suffices. The computational cost of LU decomposition is approximately proportional to the square of the number of points in the SEM. Consequently, from a cost standpoint, there is a trade-off between having more elements and a lower order scheme, and fewer elements with a higher order scheme. We have found that 5 point SEM is, in general, a reasonable middle-ground. 

\subsection{The outer core}
The discussion thus far has focused on benchmarking the code against other codes within the seismic regime. We have seen that the outer core undertones are not appreciably excited. However, challenges associated with fluid regions at low frequencies still exist. To demonstrate this we consider a tidal forcing. The tidal forcing potential has the radial dependence $\Psi_{lm} \propto r^l$~\citep{wahr1981body}.

\begin{figure*}
  \centering
    \includegraphics[width=.7\textwidth, trim = {3.3cm 4cm 2cm 3cm},clip]{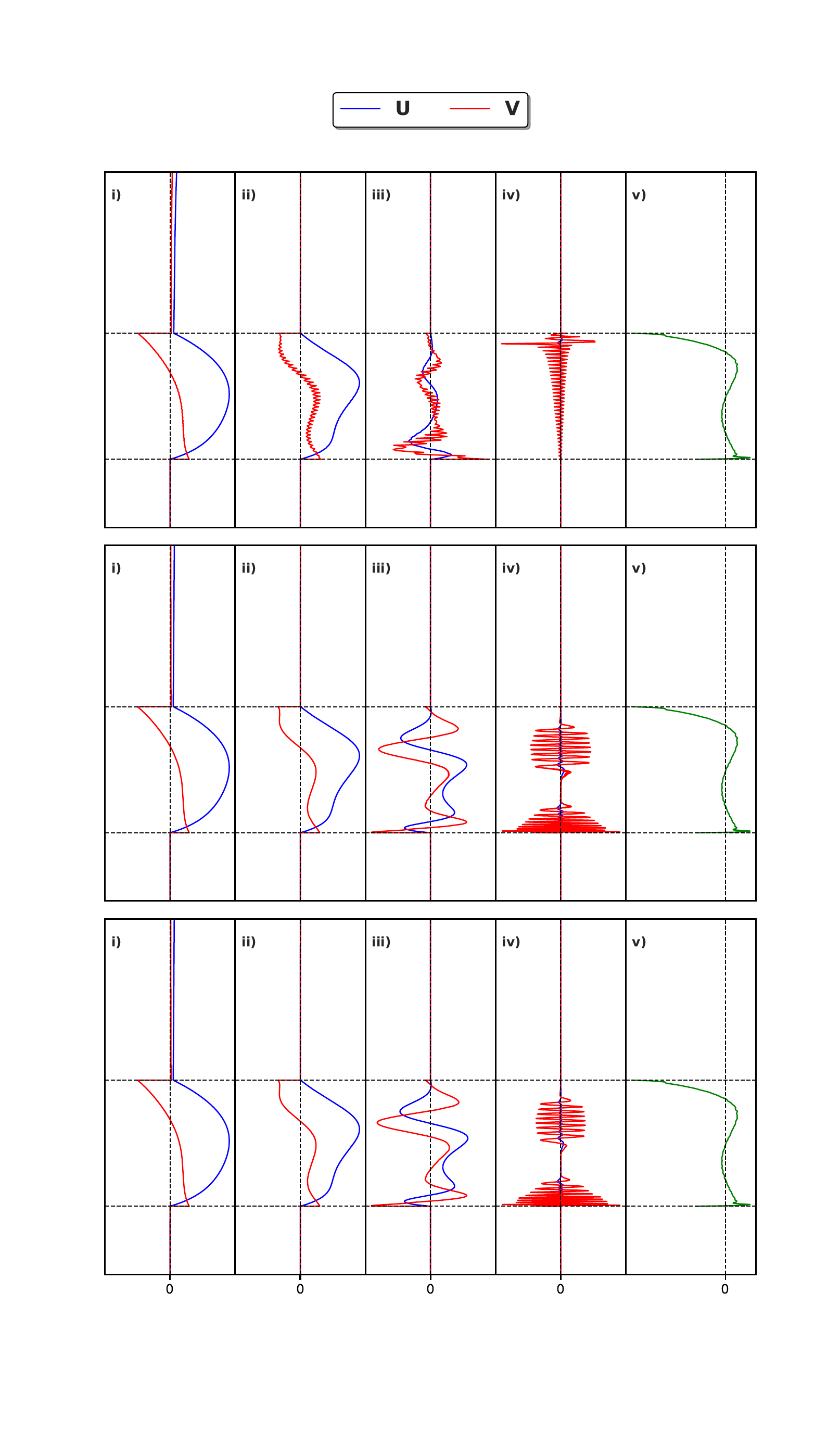}
    \caption{Example of excitation of extraneous modes at low frequencies. The maximum element size chosen was $0.01a$, $0.001a$ and $0.0001a$ for the three panels respectively (where $a$ is the Earth's radius). The period at which the linear system was solved was $(1.3, 5.2, 18.2, 182.2)$ respectively for plots i) through iv), expressed in terms of the maximum Brunt-V\"ais\"al\"a period (which is 9.25h). Plot v) shows the squared Brunt-V\"ais\"al\"a frequency.}
  \label{fig:tides_excitation}
\end{figure*}

In Fig.~\ref{fig:tides_excitation} we demonstrate the excitement of the undertones within the outer core for the degree two tide, as the frequency trends towards zero. We note the period at which we excite the system by its ratio to the maximum Brunt-V\"ais\"al\"a (BV) period (approximately 9.25 h). As the period at which we excite the system increases to greater than the BV period, the excitation within the outer core increases markedly. It is noteworthy that at relatively low resolutions the response is clearly unphysical, as in plot ii through iv in the upper panel of Fig.~\ref{fig:tides_excitation}. However, at a sufficiently small mesh size the response becomes well resolved, as in plot iv of the bottom panel in Fig.~\ref{fig:tides_excitation}. This suggests that for a particular mesh there is both a minimum and maximum frequency at which it resolves the solution, if there is appreciable excitation within the fluid outer core. The situation within laterally heterogeneous and rotating Earth models is, however, likely to be more complicated~\citep{valette1989spectre}.

\section{Conclusions}
In this paper we have discussed and benchmarked our new code, \texttt{DSpecM1D}, for the calculation of seismograms in SNRATI models. We have significantly extended the direct solution method by using radial spectral elements. In addition, we have solved the equations of motion with the inclusion of self-gravity. By using the same representation of the displacement in fluid regions we can also apply \texttt{DSpecM1D} to models with non neutrally-stratified outer cores. 

Beyond its immediate application to calculations in spherically symmetric Earth models, this code has been designed to provide the necessary ``spherical Earth preconditioner'' within the iterative solution of the equations of motion in rotating and laterally heterogeneous Earth models, following the method developed by \cite{maitra2019non} and \cite{myhill2025forward} for solving Poisson's equation within aspherical planets.

\begin{acknowledgments}
AM wishes to acknowledge the support of a NERC DTP studentship as well as a Gates Cambridge scholarship. 
\end{acknowledgments}

\paragraph*{DATA AVAILABILITY\\}
A first version of the library DSpecM1D is available on GitHub at https://github.com/adcm2/DSpecM1D.git.

\bibliographystyle{gji}
\bibliography{bibliography}

@book{Dahlen_Tromp_1998,
  url         = {https://doi.org/10.1515/9780691216157},
  title       = {Theoretical Global Seismology},
  author      = {F. A. Dahlen and Jeroen Tromp},
  publisher   = {Princeton University Press},
  address     = {Princeton},
  doi         = {doi:10.1515/9780691216157},
  isbn        = {9780691216157},
  year        = {1998},
  lastchecked = {2023-10-05}
}

@article{ishii1999normal,
  title     = {Normal-mode and free-air gravity constraints on lateral variations in velocity and density of Earth's mantle},
  author    = {Ishii, Miaki and Tromp, Jeroen},
  journal   = {Science},
  volume    = {285},
  number    = {5431},
  pages     = {1231--1236},
  year      = {1999},
  publisher = {American Association for the Advancement of Science}
}

@article{maitra2019non,
  title     = {A non-perturbative method for gravitational potential calculations within heterogeneous and aspherical planets},
  author    = {Maitra, Matthew and Al-Attar, David},
  journal   = {Geophysical Journal International},
  volume    = {219},
  number    = {2},
  pages     = {1043--1055},
  year      = {2019},
  publisher = {Oxford University Press}
}

@article{lau2017tidal,
  title     = {Tidal tomography constrains Earth’s deep-mantle buoyancy},
  author    = {Lau, Harriet CP and Mitrovica, Jerry X and Davis, James L and Tromp, Jeroen and Yang, Hsin-Ying and Al-Attar, David},
  journal   = {Nature},
  volume    = {551},
  number    = {7680},
  pages     = {321--326},
  year      = {2017},
  publisher = {Nature Publishing Group UK London}
}

@article{akbarashrafi2018exact,
  title     = {Exact free oscillation spectra, splitting functions and the resolvability of Earth's density structure},
  author    = {Akbarashrafi, F and Al-Attar, David and Deuss, Arwen and Trampert, Jeannot and Valentine, AP},
  journal   = {Geophysical Journal International},
  volume    = {213},
  number    = {1},
  pages     = {58--76},
  year      = {2018},
  publisher = {Oxford University Press}
}

@article{al2018hamilton,
  title     = {Hamilton’s principle and normal mode coupling in an aspherical planet with a fluid core},
  author    = {Al-Attar, David and Crawford, Ophelia and Valentine, Andrew P and Trampert, Jeannot},
  journal   = {Geophysical Journal International},
  volume    = {214},
  number    = {1},
  pages     = {485--507},
  year      = {2018},
  publisher = {Oxford University Press}
}

@article{al2012calculation,
  title     = {Calculation of normal mode spectra in laterally heterogeneous earth models using an iterative direct solution method},
  author    = {Al-Attar, David and Woodhouse, John H and Deuss, Arwen},
  journal   = {Geophysical Journal International},
  volume    = {189},
  number    = {2},
  pages     = {1038--1046},
  year      = {2012},
  publisher = {Blackwell Publishing Ltd Oxford, UK}
}

@article{gharti2023spectral,
  title     = {Spectral-infinite-element simulations of seismic wave propagation in self-gravitating, rotating 3-D Earth models},
  author    = {Gharti, Hom Nath and Eaton, Will and Tromp, Jeroen},
  journal   = {Geophysical Journal International},
  volume    = {235},
  number    = {3},
  pages     = {2671--2693},
  year      = {2023},
  publisher = {Oxford University Press}
}

@inproceedings{geller1990new,
  title     = {A new algorithm for waveform inversion using a laterally heterogeneous starting model},
  author    = {Geller, RJ},
  booktitle = {Prog. Abstr. Seismol. Soc. Jpn},
  volume    = {2},
  pages     = {296},
  year      = {1990}
}

@article{geller1994computation,
  title     = {Computation of synthetic seismograms and their partial derivatives for heterogeneous media with arbitrary natural boundary conditions using the Direct Solution Method},
  author    = {Geller, Robert J and Ohminato, Takao},
  journal   = {Geophysical Journal International},
  volume    = {116},
  number    = {2},
  pages     = {421--446},
  year      = {1994},
  publisher = {Blackwell Publishing Ltd Oxford, UK}
}

@article{cummins1994dsm,
  title     = {DSM complete synthetic seismograms: SH, spherically symmetric, case},
  author    = {Cummins, Phil R and Geller, Robert J and Hatori, Tomohiko and Takeuchi, Nozomu},
  journal   = {Geophysical research letters},
  volume    = {21},
  number    = {7},
  pages     = {533--536},
  year      = {1994},
  publisher = {Wiley Online Library}
}

@article{cummins1994dsm2,
  title     = {DSM complete synthetic seismograms: P-SV, spherically symmetric, case},
  author    = {Cummins, Phil R and Geller, Robert J and Takeuchi, Nozomu},
  journal   = {Geophysical research letters},
  volume    = {21},
  number    = {15},
  pages     = {1663--1666},
  year      = {1994},
  publisher = {Wiley Online Library}
}

@article{cummins1997computation,
  title     = {Computation of complete synthetic seismograms for laterally heterogeneous models using the Direct Solution Method},
  author    = {Cummins, Phil R and Takeuchi, Nozomu and Geller, Robert J},
  journal   = {Geophysical Journal International},
  volume    = {130},
  number    = {1},
  pages     = {1--16},
  year      = {1997},
  publisher = {Blackwell Publishing Ltd Oxford, UK}
}

@article{kawai2006complete,
  title     = {Complete synthetic seismograms up to 2 Hz for transversely isotropic spherically symmetric media},
  author    = {Kawai, Kenji and Takeuchi, Nozomu and Geller, Robert J},
  journal   = {Geophysical Journal International},
  volume    = {164},
  number    = {2},
  pages     = {411--424},
  year      = {2006},
  publisher = {Blackwell Publishing Ltd Oxford, UK}
}

@article{kemper2022spectral,
  title     = {A spectral element approach to computing normal modes},
  author    = {Kemper, Johannes and van Driel, Martin and Munch, F and Khan, Amir and Giardini, Domenico},
  journal   = {Geophysical Journal International},
  volume    = {229},
  number    = {2},
  pages     = {915--932},
  year      = {2022},
  publisher = {Oxford University Press}
}

@article{komatitsch1998spectral,
  title     = {The spectral element method: an efficient tool to simulate the seismic response of 2D and 3D geological structures},
  author    = {Komatitsch, Dimitri and Vilotte, Jean-Pierre},
  journal   = {Bulletin of the seismological society of America},
  volume    = {88},
  number    = {2},
  pages     = {368--392},
  year      = {1998},
  publisher = {The Seismological Society of America}
}

@article{nissen2014axisem,
  title     = {AxiSEM: broadband 3-D seismic wavefields in axisymmetric media},
  author    = {Nissen-Meyer, Tarje and van Driel, Martin and St{\"a}hler, Simon C and Hosseini, Kasra and Hempel, Stefanie and Auer, Ludwig and Colombi, Andrea and Fournier, Alexandre},
  journal   = {Solid Earth},
  volume    = {5},
  number    = {1},
  pages     = {425--445},
  year      = {2014},
  publisher = {Copernicus GmbH}
}

@article{buland1984computation,
  title     = {Computation of free oscillations of the Earth},
  author    = {Buland, Ray and Gilbert, Freeman},
  journal   = {Journal of Computational Physics},
  volume    = {54},
  number    = {1},
  pages     = {95--114},
  year      = {1984},
  publisher = {Elsevier}
}

@article{forbriger2005proposal,
  title     = {A proposal for a consistent parametrization of earth models},
  author    = {Forbriger, Thomas and Friederich, Wolfgang},
  journal   = {Geophysical Journal International},
  volume    = {162},
  number    = {2},
  pages     = {425--430},
  year      = {2005},
  publisher = {Blackwell Publishing Ltd Oxford, UK}
}

@article{al2008calculation,
  title     = {Calculation of seismic displacement fields in self-gravitating earth models—applications of minors vectors and symplectic structure},
  author    = {Al-Attar, David and Woodhouse, John H},
  journal   = {Geophysical Journal International},
  volume    = {175},
  number    = {3},
  pages     = {1176--1208},
  year      = {2008},
  publisher = {Blackwell Publishing Ltd Oxford, UK}
}

@misc{mineos,
  author    = {Masters, G. and Woodhouse, J. H. and Freeman, G.},
  title     = {Mineos v1.0.2 [software]},
  year      = {2011},
  publisher = {Computational Infrastructure for Geodynamics},
  url       = {https://geodynamics.org}
}

@misc{eigenweb,
  author       = {Ga\"{e}l Guennebaud and Beno\^{i}t Jacob and others},
  title        = {Eigen},
  howpublished = {https://libeigen.gitlab.io},
  year         = {2010}
}

@article{dziewonski1981preliminary,
  title   = {Preliminary reference Earth model},
  journal = {Physics of the Earth and Planetary Interiors},
  volume  = {25},
  number  = {4},
  pages   = {297-356},
  year    = {1981},
  issn    = {0031-9201},
  doi     = {https://doi.org/10.1016/0031-9201(81)90046-7},
  url     = {https://www.sciencedirect.com/science/article/pii/0031920181900467},
  author  = {Dziewonski, Adam M and Anderson, Don L. }
}

@book{igel2017computational,
  title     = {Computational seismology: a practical introduction},
  author    = {Igel, Heiner},
  year      = {2017},
  publisher = {Oxford University Press}
}

@article{leng2019axisem3d,
  title     = {AxiSEM3D: broad-band seismic wavefields in 3-D global earth models with undulating discontinuities},
  author    = {Leng, Kuangdai and Nissen-Meyer, Tarje and Van Driel, Martin and Hosseini, Kasra and Al-Attar, David},
  journal   = {Geophysical Journal International},
  volume    = {217},
  number    = {3},
  pages     = {2125--2146},
  year      = {2019},
  publisher = {Oxford University Press}
}

@article{valette1989spectre,
  title   = {Spectre des vibrations propres d’un corps {\'e}lastique, auto-gravitant, en rotation uniforme et contenant une partie fluide},
  author  = {Valette, B},
  journal = {CR Acad. Sci. Paris},
  volume  = {309},
  number  = {S{\'e}rie I},
  pages   = {419--422},
  year    = {1989}
}

@article{phinney1973representation,
  title     = {Representation of the elastic-gravitational excitation of a spherical Earth model by generalized spherical harmonics},
  author    = {Phinney, Robert A and Burridge, Robert},
  journal   = {Geophysical Journal International},
  volume    = {34},
  number    = {4},
  pages     = {451--487},
  year      = {1973},
  publisher = {Blackwell Publishing Ltd Oxford, UK}
}

@article{al2014sensitivity,
  title     = {Sensitivity kernels for viscoelastic loading based on adjoint methods},
  author    = {Al-Attar, David and Tromp, Jeroen},
  journal   = {Geophysical Journal International},
  volume    = {196},
  number    = {1},
  pages     = {34--77},
  year      = {2014},
  publisher = {Oxford University Press}
}

@article{crotwell1999taup,
  title     = {The TauP Toolkit: Flexible seismic travel-time and ray-path utilities},
  author    = {Crotwell, H Philip and Owens, Thomas J and Ritsema, Jeroen and others},
  journal   = {Seismological Research Letters},
  volume    = {70},
  pages     = {154--160},
  year      = {1999},
  publisher = {Seismological Society of America}
}

@software{crotwell_2025_16884103,
  author    = {Crotwell, H. Philip},
  title     = {The TauP Toolkit},
  month     = aug,
  year      = 2025,
  publisher = {Zenodo},
  version   = {3.1.0},
  doi       = {10.5281/zenodo.16884103},
  url       = {https://doi.org/10.5281/zenodo.16884103}
}

@article{myhill2025forward,
  title   = {Forward and adjoint calculations of gravitational potential in heterogeneous, aspherical planets},
  author  = {Myhill, Alex DC and Maitra, Matthew A and Al-Attar, David},
  journal = {arXiv preprint arXiv:2508.07910},
  year    = {2025}
}

@article{chaljub2004spectral,
  title={Spectral element modelling of three-dimensional wave propagation in a self-gravitating Earth with an arbitrarily stratified outer core},
  author={Chaljub, Emmanuel and Valette, Bernard},
  journal={Geophysical Journal International},
  volume={158},
  number={1},
  pages={131--141},
  year={2004},
  publisher={Blackwell Publishing Ltd Oxford, UK}
}

@article{wahr1981body,
  title={Body tides on an elliptical, rotating, elastic and oceanless Earth},
  author={Wahr, John M},
  journal={Geophysical Journal International},
  volume={64},
  number={3},
  pages={677--703},
  year={1981},
  publisher={Blackwell Publishing Ltd Oxford, UK}
}

@article{diaz1977collocation,   
  title={A collocation-Galerkin method for the two point boundary value problem using continuous piecewise polynomial spaces},   
  author={Diaz, Julio C},   
  journal={SIAM Journal on Numerical Analysis},   
  volume={14},   
  number={5},   
  pages={844--858},   
  year={1977},   
  publisher={SIAM} 
}

@article{al2016particle,
  title={Particle relabelling transformations in elastodynamics},
  author={Al-Attar, David and Crawford, Ophelia},
  journal={Geophysical Supplements to the Monthly Notices of the Royal Astronomical Society},
  volume={205},
  number={1},
  pages={575--593},
  year={2016},
  publisher={The Royal Astronomical Society}
}

@article{deuss2001theoretical,
  title={Theoretical free-oscillation spectra: the importance of wide band coupling},
  author={Deuss, Arwen and Woodhouse, John H},
  journal={Geophysical Journal International},
  volume={146},
  number={3},
  pages={833--842},
  year={2001},
  publisher={Blackwell Publishing Ltd Oxford, UK}
}

@article{yang2015synthetic,
  title={Synthetic free-oscillation spectra: an appraisal of various mode-coupling methods},
  author={Yang, Hsin-Ying and Tromp, Jeroen},
  journal={Geophysical Journal International},
  volume={203},
  number={2},
  pages={1179--1192},
  year={2015},
  publisher={Oxford University Press}
}

@article{woodhouse1980coupling,
  title={The coupling and attenuation of nearly resonant multiplets in the Earth's free oscillation spectrum},
  author={Woodhouse, JH},
  journal={Geophysical Journal International},
  volume={61},
  number={2},
  pages={261--283},
  year={1980},
  publisher={Blackwell Publishing Ltd Oxford, UK}
}

@article{woodhouse1976rayleigh,
  title={On Rayleigh's principle},
  author={Woodhouse, JH},
  journal={Geophysical Journal International},
  volume={46},
  number={1},
  pages={11--22},
  year={1976},
  publisher={Blackwell Publishing Ltd Oxford, UK}
}

@article{van2021modelling,
  title={On the modelling of self-gravitation for full 3-D global seismic wave propagation},
  author={van Driel, Martin and Kemper, Johannes and Boehm, Christian},
  journal={Geophysical Journal International},
  volume={227},
  number={1},
  pages={632--643},
  year={2021},
  publisher={Oxford University Press}
}

@article{adourian2024adjoint,
  title={Adjoint sensitivity kernels for free oscillation spectra},
  author={Adourian, S and Dursun, MS and Lau, HCP and Al-Attar, D},
  journal={Geophysical Journal International},
  volume={238},
  number={1},
  pages={257--271},
  year={2024},
  publisher={Oxford University Press}
}

@article{leng2016efficient,
  title={Efficient global wave propagation adapted to 3-D structural complexity: a pseudospectral/spectral-element approach},
  author={Leng, Kuangdai and Nissen-Meyer, Tarje and van Driel, Martin},
  journal={Geophysical Supplements to the Monthly Notices of the Royal Astronomical Society},
  volume={207},
  number={3},
  pages={1700--1721},
  year={2016},
  publisher={The Royal Astronomical Society}
}

@article{davies2015thermally,
  title={Thermally dominated deep mantle LLSVPs: a review},
  author={Davies, DR and Goes, S and Lau, HCP},
  journal={The Earth's heterogeneous mantle: A geophysical, geodynamical, and geochemical perspective},
  pages={441--477},
  year={2015},
  publisher={Springer}
}

@article{valette1986influence,
  title={About the influence of pre-stress upon adiabatic perturbations of the Earth},
  author={Valette, Bernard},
  journal={Geophysical Journal International},
  volume={85},
  number={1},
  pages={179--208},
  year={1986},
  publisher={Blackwell Publishing Ltd Oxford, UK}
}

@article{woodhouse1978effect,
  title={The effect of a general aspherical perturbation on the free oscillations of the Earth},
  author={Woodhouse, Jim H and Dahlen, Francis A},
  journal={Geophysical Journal International},
  volume={53},
  number={2},
  pages={335--354},
  year={1978},
  publisher={Blackwell Publishing Ltd Oxford, UK}
}

@article{helffrich2013causes,
  title={Causes and consequences of outer core stratification},
  author={Helffrich, George and Kaneshima, Satoshi},
  journal={Physics of the Earth and Planetary Interiors},
  volume={223},
  pages={2--7},
  year={2013},
  publisher={Elsevier}
}

@book{schechter2001principles,
  title={Principles of functional analysis},
  author={Schechter, Martin},
  number={36},
  year={2001},
  publisher={American Mathematical Soc.}
}

\appendix
\section{Equation of motion in an SNRATI Earth model}
In this appendix we discuss the reduction of the equations of motion in an SNRATI Earth to the toroidal and spheroidal systems of ODEs. We firstly introduce the generalised spherical harmonics, and the representation of the displacement vector. The forcing terms and toroidal and spheroidal equations are then discussed. 

\subsection{Generalised spherical harmonics}
We recall the definitions of the canonical basis vectors of~\citet{phinney1973representation}
\begin{align}
  \hat{\mathbf{e}}_{-} & = \frac{1}{\sqrt{2}} (\hat{\boldsymbol{\theta}} - i \hat{\boldsymbol{\phi}}), \\
  \hat{\mathbf{e}}_0 & = \hat{\mathbf{r}}, \\
  \hat{\mathbf{e}}_{+} & = -\frac{1}{\sqrt{2}} (\hat{\boldsymbol{\theta}} + i \hat{\boldsymbol{\phi}}).
\end{align}
Any vector field can be represented as 
\begin{equation}
  \mathbf{u} =  \sum_{lm} u_{lm}^{\alpha} Y_{lm}^\alpha\hat{\mathbf{e}}_{\alpha},
\end{equation}
where $\alpha = \{-,0,+\}$, $Y_{lm}^{\alpha}$ are the normalised generalised spherical harmonics (GSH) of Appendix C of~\citet{Dahlen_Tromp_1998}, summation is over integer values for $0\leq l \leq \infty$ and $-l \leq m \leq l$ and summation over GSH upper indices $\alpha$ is implied. We define the set of coefficients $U,V,W$ as 
\begin{align}
  u_{lm}^{\pm} & = \frac{1}{\sqrt{2}} \zeta (V_{lm} \pm i W_{lm}), \\
  u_{lm}^0 & = U_{lm},
\end{align}
where $\zeta = \sqrt{l(l + 1)}$. The coefficients $W_{lm}$ are the toroidal components of the wavefield whilst $U_{lm}$ and $V_{lm}$ are the spheroidal components. The ground motion in spherical polar coordinates is given by
\begin{align}
  u_r & = \sum_{lm} U_{lm} Y_{lm}^0, \\
  u_{\theta} & = \sum_{lm} \frac{\zeta}{2} \left[(Y_{lm}^- - Y_{lm}^+)V_{lm} - i (Y_{lm}^- + Y_{lm}^+) W_{lm}\right], \\
  u_{\varphi} & = \sum_{lm} \frac{i\zeta}{2} \left[(Y_{lm}^- + Y_{lm}^+)V_{lm} - i (Y_{lm}^- - Y_{lm}^+) W_{lm}\right].
\end{align}
We also expand the gravitational potential perturbation associated with spheroidal motion (toroidal motion does not change the gravitational potential) using spherical harmonics, i.e.
\begin{equation}
   \phi = \sum_{lm} \phi_{lm}Y_{lm}^0.
\end{equation}

The reduction of the equation of motion for a fully isotropic model, using the generalised spherical harmonic representation, is detailed in Appendix B of~\citet{al2014sensitivity}. We use a similar method to reduce the equations of motion to the appropriate radial integrals, in the case of transverse isotropy. In the interest of brevity we do not detail the reduction but state the results in subsequent sub-appendices. Finally, we note that the equation we obtain for the spheroidal component differs to that found by~\citet{kemper2022spectral} in the presence of a negative term in front of the integral involving the Love parameter $F$. This difference is due to a mistake within their manuscript, as can be seen by the fact that their equations do not reduce in the isotropic case to the expressions detailed in \cite{al2014sensitivity}.

\subsection{Excitation}
The two types of forcing we consider in this paper are: a moment-tensor point source and tidal forcing. Tidal forcing is used to illustrate the effect of undertones in the outer core. The tidal potential is given by
\begin{equation}
  \Psi = \sum_{l=2}^{\infty} \sum_{m = 0}^l \left(\frac{r}{a}\right)^l c_{lm} Y_{lm},
\end{equation}
where $c_{lm}$ depends on the position of the sun and moon~\citep{wahr1981body}. Furthermore, the degree two and three terms dominate. We do not elaborate further as tidal forcing is used for illustrative purposes only. 

Moment-tensor point sources are used within seismology to represent the forcing associated with an Earthquake. The force operator (in the time domain) that corresponds to a moment-tensor point source is
\begin{equation}
  \mathcal{F}(\mathbf{u}') = \mathbf{M} : \overline{(\boldsymbol{\nabla}\mathbf{u}')(\mathbf{r}_s)} m(t),
\end{equation}
where $\mathbf{M}$ is the moment tensor for the Earthquake, $\mathbf{u}'$ is the test function, $m(t)$ is the source-time function (STF), $\mathbf{r}_s$ is the source location and an overbar indicates taking a complex conjugate. We use a Heaviside step-function for $m(t)$. If one wishes to use another STF, the seismogram corresponding to a Heaviside step-function can simply be 
convolved against the derivative of the STF that one is using. 

\subsection{Toroidal equation}
The toroidal subsystem includes only the coefficient $W_{lm}$. As the outer core is fluid, it cannot accommodate shear. Consequently, there are two sets of toroidal motions: those within the inner core and those within the mantle. As inner core motions are not excited or observed on the surface we neglect them herein. The equations for different degrees decouple entirely and the inertia and stiffness operators are independent of order. This means that we can independently define the inertia and stiffness operators at different degrees. These are given by
\begin{equation}
  \mathcal{P}_l(W', W) = \zeta^2 \int_{b}^{a} \rho \overline{W'} W r^2 \dd r,
\end{equation}
and 
\begin{align}
  & \mathcal{H}_l(\omega,W', W)  \nonumber \\
  & = \zeta^2\int_b^a L(\omega) (r\partial_r \overline{W'} - \overline{W'})(r\partial_r W - W) \dd r \nonumber \\
  & \phantom{=}+  \zeta^2 \chi^2\int_{b}^{a} N(\omega) \overline{W'} W \dd r,
\end{align}
where $b$ is the radius of the core-mantle boundary, $a$ is the Earth's radius, $\chi^2 = \zeta^2 - 2$ and we have dropped the $lm$ subscript. 
The corresponding force term is given by (excluding the source-time function)
\begin{align}
  & \frac{1}{2}\zeta r_s^{-1}\bigg\{r_s (\partial_r \overline{W'})(r_s) \left[M_{r\theta} \overline{S_+^1(\theta_s,\varphi_s)} + iM_{r\varphi}\overline{S_-^1(\theta_s,\varphi_s)}\right] \nonumber\\
  & + \overline{W'}(r_s)\bigg[\frac{1}{2} \chi(M_{\theta\theta} - M_{\varphi\varphi})\overline{S_-^2(\theta_s,\varphi_s)} - M_{r\theta} \overline{S_+^1(\theta_s,\varphi_s)} \nonumber\\
  &  + i \left(\chi M_{\theta \varphi}\overline{S_+^2(\theta_s,\varphi_s)}-M_{r\varphi}\overline{S_-^1(\theta_s,\varphi_s)}\right)  \bigg]\bigg\},
\end{align}
where we have defined $S_{\pm}^\alpha(\theta,\varphi) = Y_{lm}^{-\alpha}(\theta,\varphi)  \pm Y_{lm}^{\alpha}(\theta,\varphi) $, 
$\mathbf{M}$ is the moment tensor for the Earthquake, and the Earthquake is located at $(r_s,\theta_s,\varphi_s)$ in spherical coordinates.

\subsection{Spheroidal equations}
The inertia operator for the one-dimensional spheroidal equation is 
\begin{align}
    \mathcal{P}_l(U',V',\phi',U,V,\phi) & = \int_{0}^{a}  \rho r^2 (\overline{U'} U + \zeta^2 \overline{V'}V) \dd r, 
\end{align}
whilst the stiffness operator is given by
\begin{align}
    & \mathcal{H}_l(\omega,U',V',\phi',U,V,\phi) \nonumber \\
    & =  \zeta^2 (\zeta^2 - 2) \int_{0}^{a} N \overline{V'}  V \dd r  + \int_{0}^{a} C \overline{\partial_r U'} \partial_r U r^2 \dd r \nonumber\\
    & \phantom{=}+ 4 \int_{0}^{a} \rho(\pi G \rho r - g) r \overline{U'} U \dd r \nonumber\\
    & \phantom{=}+ \zeta^2 \int_{0}^{a} L [\overline{\partial_r V'} - r^{-1} (\overline{V'} - \overline{U'})][\partial_{r}V - r^{-1} (V - U)] r^2 \dd r \nonumber\\
    & \phantom{=}+ \int_{0}^{a} (A- N) (2 \overline{U'} - \zeta^2 \overline{V'}) (2 U - \zeta^2 V) \dd r \nonumber\\
    & \phantom{=}+ \int_{0}^{a} F \left[2 (\overline{\partial_r U'} U + \overline{U'} \partial_{r}U) - \zeta^2 (\overline{\partial_r U'} V + \partial_{r}U \overline{V'} )\right]r \dd r \nonumber\\
    & \phantom{=}+ \frac{1}{4\pi G}\int_{0}^{a} \left[r^2 \overline{\partial_r \phi'}\partial_r\phi + \zeta^2 \overline{\phi'}\phi\right]\dd r + \frac{1}{4\pi G}a(l + 1) \overline{\phi'}\phi\nonumber\\
    & \phantom{=}+ \int_{0}^{a} \rho [\overline{\partial_r \phi'} U + \partial_r \phi \overline{U'}] r^2 \dd r + \zeta^2 \int_{0}^{a} \rho [\overline{\partial_r \phi'} V + \phi \overline{V'}] r \nonumber \\
    & \phantom{=}+ \zeta^2 \int_{0}^{a} \rho g [\overline{U'} V + U \overline{V'}] r \dd r.
\end{align}
The force term is given by
\begin{align}
  & M_{rr} \overline{(\partial_r U')(r_s) Y_{lm}^0} \nonumber\\
  & + r_s^{-1} \overline{U'(r_s)} \Big\{\frac{\zeta}{2}\big[iM_{r\varphi}\overline{S_+^1(\theta_s,\varphi_s)} + M_{r\theta} \overline{S_-^1(\theta_s,\varphi_s)}\big] \nonumber\\
  & + (M_{\theta \theta} + M_{\varphi\varphi})\overline{Y_{lm}^0(\theta_s,\varphi_s)}\Big\} \nonumber\\
  & + \frac{\zeta}{2} \overline{(\partial_r V')(r_s)} \big[iM_{r\varphi}\overline{S_+^1(\theta_s,\varphi_s)} + M_{r\theta} \overline{S_-^1(\theta_s,\varphi_s)}\big] \nonumber\\
  & + \frac{\zeta}{2} r_s^{-1} \overline{V'(r_s)} \bigg\{-\zeta(M_{\theta \theta} + M_{\varphi\varphi})\overline{Y_{lm}^0(\theta_s,\varphi_s)}\nonumber\\
  & -\big[iM_{r\varphi}\overline{S_+^1(\theta_s,\varphi_s)} + M_{r\theta} \overline{S_-^1(\theta_s,\varphi_s)}\big] \nonumber\\
  & + \chi \Big[\frac{1}{2}(M_{\theta \theta} - M_{\varphi\varphi})\overline{S_+^2(\theta_s,\varphi_s)} + i M_{\theta\varphi}\overline{S_-^2(\theta_s,\varphi_s)}\Big]\bigg\}.
\end{align}

\subsection{Constant-Q dispersion model}
Dispersive models possess frequency dependent Love parameters. We consider the constant-Q dispersion model as a specific example. Although it is possible to specify the frequency dependence of each Love parameter, most spherically symmetric Earth models tend to specify only the frequency dependence of the equivalent bulk and shear moduli, i.e.
\begin{align}
\mu(\omega) &= \mu_0 \left[1 + \frac{2}{\pi Q_\mu} \ln\left(\frac{i\omega}{\omega_0}\right) \right], \label{eqn:qdisp_mu}\\
\kappa(\omega) &= \kappa_0 \left[1 + \frac{2}{\pi Q_\kappa} \ln\left(\frac{i\omega}{\omega_0}\right)\right], \label{eqn:qdisp_kappa}
\end{align}
where $\mu_0 = \mu(\omega_0)$ and $\kappa_0 = \kappa(\omega_0)$ are the shear and bulk moduli evaluated at the reference frequency $\omega_0$ whilst $Q_\mu$ and $Q_\kappa$ are the shear and bulk quality factors. In the case of a transversely isotropic Earth model we take the approach of Chapter 9.7 of~\citet{Dahlen_Tromp_1998}, which is also the suggested approach of~\citet{forbriger2005proposal}, and has long been applied within codes such as \texttt{MINEOS} and \texttt{YSpec}. The Love parameters are given by
\begin{align}
    A & = \delta A + \kappa + \frac{4}{3}\mu, \\
    C & = \delta C + \kappa + \frac{4}{3}\mu, \\
    L & = \delta L + \mu, \\
    N & = \delta N + \mu, \\
    F & = \delta F + \kappa - \frac{2}{3} \mu, 
\end{align}
where 
\begin{align}
  \kappa & = \frac{1}{9}(C + 4A - 4N + 4F), \\
  \mu & = \frac{1}{15}(C + A + 6L + 5N - 2F),
\end{align}
are the ``equivalent'' isotropic bulk and shear moduli, corresponding to the ``closest'', in the least squares sense, isotropic model to the transversely isotropic one~\citep{Dahlen_Tromp_1998}. The Love parameters are the sum of ``purely anisotropic'' (indicated by the prefix $\delta$) and isotropic components. The ``purely anisotropic'' components are assumed to be frequency independent, whilst the equivalent isotropic moduli $\kappa$ and $\mu$ follow eqs~\eqref{eqn:qdisp_mu} and~\eqref{eqn:qdisp_kappa}. This implies that
\begin{align}
    A(\omega) & = A_0 + \frac{2}{\pi}\left[\frac{\kappa_0}{ Q_{\kappa}} + \frac{4\mu_0}{3Q_{\mu}}\right] \ln \left(\frac{i\omega}{\omega_0}\right), \\
    C(\omega) & = C_0 + \frac{2}{\pi}\left[\frac{\kappa_0}{ Q_{\kappa}} + \frac{4\mu_0}{3Q_{\mu}}\right] \ln \left(\frac{i\omega}{\omega_0}\right), \\
    L(\omega) & = L_0 +\frac{2\mu_0}{\pi Q_{\mu}} \ln \left(\frac{i\omega}{\omega_0}\right), \\
    N(\omega) & = N_0 +\frac{2\mu_0}{\pi Q_{\mu}} \ln \left(\frac{i\omega}{\omega_0}\right), \\
    F(\omega) & = F_0 + \frac{2}{\pi}\left[\frac{\kappa_0}{ Q_{\kappa}} - \frac{2\mu_0}{3Q_{\mu}}\right] \ln \left(\frac{i\omega}{\omega_0}\right),
\end{align}
where a subscript zero indicates evaluation at the reference frequency, e.g. $A_0 = A(\omega_0)$, etc. 

\section{Matrix assembly}
\subsection{Spectral elements}
A spectral element method subdivides the entire domain into a set of elements $\mathcal{R}_i$, where $\mathcal{R}_i \equiv [r_{i},r_{i+1}]$. Within each element a local basis is defined that has support only within that element. The basis used within each element is the set of Lagrange polynomials defined on the Gauss-Lobatto-Legendre (GLL) collocation points (quadrature nodes). This choice of interpolation points ensures that Gibbs phenomena are minimised~\citep{igel2017computational}. The Lagrange polynomials for a set of points $\{x_i\}$ are defined as
\begin{equation}
  l_i(x) = \prod_{j \neq i}^{N} \frac{x - x_j}{x_i - x_j}.
\end{equation}
The GLL points are given on the interval $\xi \in [-1,1]$, and a mapping from $[-1,1]$ to $r \in [r_i,r_{i+1}]$ must be defined. A simple linear map suffices, i.e., $r(\xi) = (r_{i + 1} - r_i) \xi/2 + (r_{i} + r_{i + 1})/2$. Radial integrals over an element are evaluated using Gauss-Lobatto-Legendre quadrature, ie 
\begin{equation}
  \int_{r_i}^{r_{i+1}} f(r) \dd r \approx \frac{r_{i + 1} - r_i}{2}\sum_{n = 0}^{N} w_n f(r_n),
\end{equation}
where $w_n$ is the $n$-th GLL weight. We represent the $j$-th basis function, in the $i$-th element by $l_{ij}(r)$.

\subsection{Dispersive models}
In a dispersive model, the inertia matrix is constant but the stiffness matrix changes with frequency, necessitating re-computing the matrix at each frequency. The cost of a naive implementation, whereby the integrals are re-evaluated at each frequency, is at least the same order of magnitude (generally greater) as the cost of solving the forced equation at each frequency\textemdash finding an efficient way to do it is imperative. In this subsection we detail the approach that we have taken when using a constant-Q logarithmic dispersion law, but it serves as a template for how this could be done more generally. The frequency dependence of the stiffness matrix is given by 
\begin{equation}
  \mathbf{H}(\omega) = \mathbf{H}(\omega_0) + \frac{2}{\pi}  \delta\mathbf{H}_0 \ln \left(\frac{i\omega}{\omega_0}\right).
\end{equation}
The matrix $\mathbf{H}(\omega_0)$ is evaluated as in the non-attenuating case. If we define the quality factor scaled parameters
\begin{align}
  A_Q & = C_Q = \left[\frac{\kappa_0}{ Q_{\kappa}} + \frac{4\mu_0}{3Q_{\mu}}\right], \\
  F_Q & = \left[\frac{\kappa_0}{ Q_{\kappa}} - \frac{2\mu_0}{3Q_{\mu}}\right], \\
  L_Q & = N_Q = \frac{\mu_0}{Q_\mu},
\end{align}
then the form corresponding to $\delta\mathbf{H}_0$ in the spheroidal case is given by 
\begin{align}
    & \mathcal{H}_0(U',V',P',U,V,P) \nonumber \\
    & =  \zeta^2 \chi^2 \int_{0}^{a} N_Q \overline{V'}  V \dd r  + \int_{0}^{a} C_Q \overline{\partial_r U'} \partial_r U r^2 \dd r \nonumber\\
    & \phantom{=}+ \zeta^2 \int_{0}^{a} L_Q [\overline{\partial_r V'} - r^{-1} (\overline{V'} - \overline{U'})][\partial_r V - r^{-1} (V - U)] r^2 \dd r \nonumber\\
    & \phantom{=}+ \int_{0}^{a} (A_Q- N_Q) (2 \overline{U'} - \zeta^2 \overline{V'}) (2 U - \zeta^2 V) \dd r \nonumber\\
    & \phantom{=}+ \int_{0}^{a} F_Q \left[2 (\overline{\partial_r U'} U + \overline{U'} \partial_r U) - \zeta^2 (\overline{\partial_r U'} V + \overline{V'} \partial_r U  )\right]r \dd r.
\end{align}
Similar forms can be defined in the toroidal and radial cases. The computational cost of matrix assembly using this technique is similar to the non-attenuating case.

\label{lastpage}
\end{document}